\documentstyle[prl,aps,amsfonts,amssymb,12pt]{revtex}
\begin{document}
\draft\tighten

\title{Spin-Waves in the Mid-Infrared Spectrum of
Antiferromagnetic YBa$_2$Cu$_3$O$_{6.0}$ }
\author{M. Gr\"uninger, J. M\"unzel, A. Gaymann, A. Zibold and
H. P.  Geserich}
\address{\it Institut f\"ur angewandte Physik, Universit\"at Karlsruhe,
D--76128 Karlsruhe, Germany}
\author{T. Kopp}
\address{Institut f\"ur Theorie der kondensierten Materie, Universit\"at
Karlsruhe}

\date{\today}
\maketitle

\begin{abstract}
The mid-infrared spin-wave spectrum of antiferromagnetic
YBa$_2$Cu$_3$O$_{6.0}$\ was determined by infrared transmission and
reflection measurements ($\bbox{k} \!\! \parallel $c) at $T\!=\!10\!$~K.\@
Excitation of single magnons of the optical branch was observed at
$E_{\text{op}}\!=\!178.0\!$~meV.\@ Two further peaks at $346\!$~meV
($\approx\!1.94\,E_{\text{op}}$) and $470\!$~meV
($\approx\!2.6\,E_{\text{op}}$)
both belong to the two-magnon spectrum. Linear spin wave theory is in good
agreement
with the measured two-magnon spectrum, and allows to determine
the exchange constant $J$  to be about $120\!$~meV,
whereas the intrabilayer coupling $J_{12}$ is approximately $0.55\,J$.
\end{abstract}

\parskip 5pt

High temperature superconductors are basically layered copper-oxide materials.
It is widely accepted that the relevant electronic degrees of freedom are
confined to
copper-oxide planes. The number of CuO$_2$ planes per unit cell varies: e.g.,
La$_{2-x}$Sr$_x$CuO$_4$ exists in a single plane form with a large spacing
between planes
of $\approx\!13.2$\AA, and  YBa$_2$Cu$_3$O$_{6+x}$ has a double layer structure
with intra- and interbilayer spacings of $\approx\!3.3$\AA\ and 8.5\AA,
respectively.
Electronic correlations, and hence spin dynamics\cite{BK}, may depend on the
type
of stacking of the planes. More specifically, a sizable coupling $J_{12}$
between spins
on adjacent planes of a bilayer will influence the spin excitation spectrum as
well
as the nature of the ground state. This may have been seen already in doped
compounds:
the normal state spin susceptibility of La$_{2-x}$Sr$_x$CuO$_4$
extrapolates to a finite value  at zero temperature, whereas it extrapolates to
zero for YBa$_2$Cu$_3$O$_{6.6}$ \cite{millis}. This may be interpreted
as a signature for the opening of a spin excitation gap in
YBa$_2$Cu$_3$O$_{6.6}$ at low
temperatures\cite{others}---a behavior certainly not encountered in Fermi
liquids.
Further, a spin density wave ordering for
La$_{2-x}$Sr$_x$CuO$_4$ has been proposed,  but for  YBa$_2$Cu$_3$O$_{6.6}$
a singlet pairing of spins in adjacent CuO$_2$ planes with strong
antiferromagnetic
fluctuations within a plane\cite{altshuler,millis,ubbens}. Such a scenario
seems to require
an unrealistically large $J_{12}\gtrsim2.5 J$\cite{sandvik}, where $J$ is the
in-plane
exchange coupling of the Heisenberg Hamiltonian supposed to describe the low
energy
spin dynamics of a single bilayer for zero doping ($x\!=\!0$).
However it was argued that, for finite doping, the itinerant carriers destroy
the
antiferromagnetism of the insulating phase and, therefore, much smaller values
of
$J_{12}$ will produce a singlet interplane pairing in the conducting phase of
YBa$_2$Cu$_3$O$_{6.6}$.

Up to now, no experimental evidence has been given of a sizable bilayer
coupling
($J_{12}\!\sim\!J$).
In neutron-scattering experiments on YBa$_2$Cu$_3$O$_{6+x}$, the in-plane
coupling
was determined from the dispersion of acoustic spin-waves and was found to be
extremely large ($J\!=\!120\pm 20\!$~meV\cite{shamoto},
$J\!=\!150\!$~meV\cite{rossat},
both for $x\!=\!0.15$). Yet, no optical modes have been found for energies up
to
60~meV\cite{shamoto,vettier}, suggesting a bilayer coupling of $J_{12}\gtrsim
8$~meV.\@
In Raman-scattering experiments on YBa$_2$Cu$_3$O$_{6+x}$
a two-magnon peak was observed\cite{lyons,sugai}. $J$  was found to be
consistent with
the neutron-scattering data, whereas $J_{12}$ was neglected.

In this Letter we report the first observation of an optical magnon peak
and of the two-magnon spectrum in infrared spectroscopy of
YBa$_2$Cu$_3$O$_{6.0}$.
They allow to determine both $J_{12}$ and $J$.
Two resonances of the two-magnon spectrum confirm the location of the
single magnon peak. Further, understanding the `antiferromagnetic limit'
YBa$_2$Cu$_3$O$_{6.0}$ will be crucial to interpret the excitations of carriers
in
doped YBa$_2$Cu$_3$O$_{6+x}$\cite{grueninger}.

Due to the high resolution and the wide spectral range, optical
spectroscopy is a powerful method to determine precisely energies of
spin-waves.
But, compared to other optical excitations such as infrared-active phonons,
intraband and interband transitions,
the absorption by magnons is two to three orders of magnitude weaker,
yielding an optical conductivity of the order of 1 S/cm.
Therefore, spin-wave excitations can be detected only in the transmittance
spectra of thin single crystals. Up to now, investigations of this type have
only been performed
on single layer cuprates\cite{perkins}.
Broad structures between 0.4~eV and 1.2~eV were observed which were
interpreted as an exciton and magnon-sidebands \cite{perkins}.

The crystals with typical dimensions of 1*1*0.1~mm$^3$ had been annealed
in the UHV at 700~K for two days to exclude doping by excess oxygen.
The samples are very close to the pure limit YBa$_2$Cu$_3$O$_{6.0}$\,
showing values of the conductivity function lower than 0.1~S/cm which is
about three (five) orders of magnitude smaller than in
YBa$_2$Cu$_3$O$_{6.1}$ (YBa$_2$Cu$_3$O$_7$) in the same spectral range.
The measurements were performed using a Fourier transform spectrometer Bruker
IFS 113v in the spectral range between 85~meV and 1.5~eV.\@
The samples were mounted on a diaphragm in a helium-flow cryostat. Reference
spectra at each temperature were obtained using a second,
identical diaphragm and a turning mechanism.
Hence an absolute photometric accuracy of the transmission data
of about 1\% was achieved.

The conductivity function
$\sigma (\hbar \omega)\!=\!2\omega \varepsilon _0 n(\hbar \omega)k(\hbar
\omega)$
can be calculated, if the sample thickness and both transmission
and reflection spectra are known. Here, $n$ and $k$ denote the real and
imaginary
part of the refractive index. The sample thickness could
be determined precisely from the spectral position of the interference
maxima. Since reflection and
transmission measurements use slightly different incident angles, and as
furthermore the incident light beams are not completely parallel, there
are still small interference structures in our plot of
$\sigma(\hbar\omega)$.

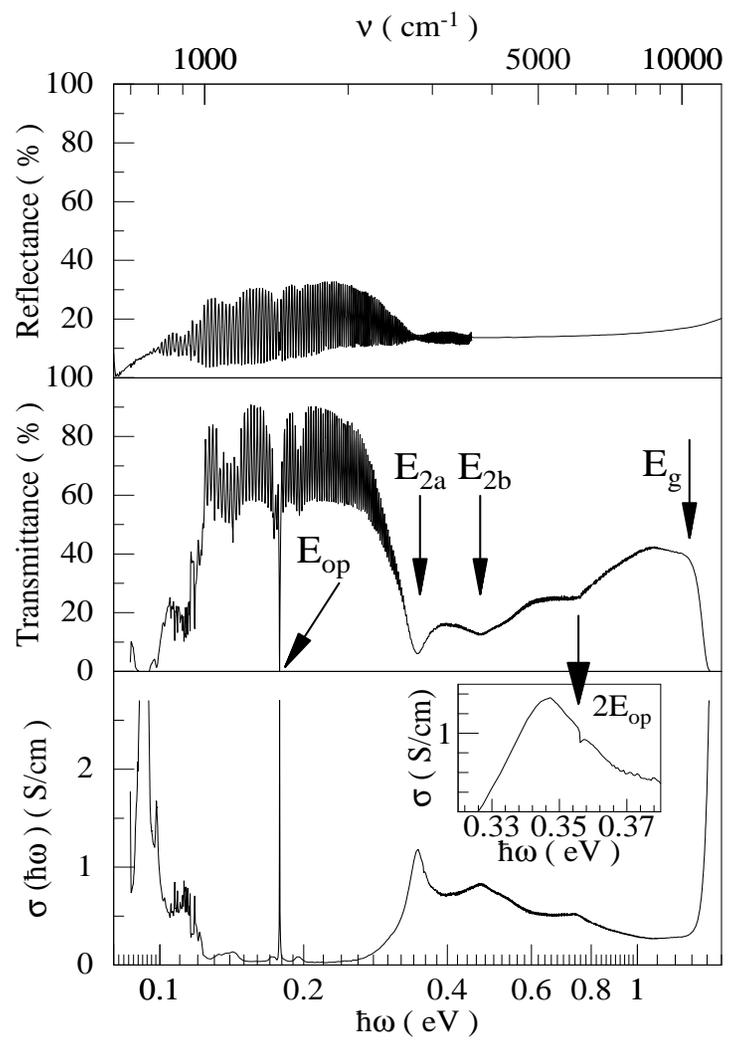
\begin{figure}
  \caption{Upper panels: In-plane reflectance and transmittance spectra
  ($\protect{\bbox{k}} \!\! \parallel $c) of a YBa$_2$Cu$_3$O$_{6.0}$\ single
crystal
  with $d\!=\!125\mu$m at $T\!=\!10\!$~K. The arrows mark absorption processes
  described in the text. Lower panel: The resulting optical conductivity
function.}
 \label{fig1}
\end{figure}

In the upper panels of Fig.\ \ref{fig1} reflectance and transmittance
spectra, obtained at $T\!=\!10\!$~K on a single-crystalline platelet with a
thickness of $d\!=\!125 \mu$m, are displayed. The resulting conductivity
function is shown in the lower panel of the same figure.
Between 0.1~eV and 0.4~eV the measurements
are dominated by interference effects, indicating regions of low absorption
(as can be seen in $\sigma $). Whereas the interference structure was
precisely resolved by the spectrometer it cannot be resolved in the
figure.

The reflectance spectrum by itself is not sufficient to determine the
excitations present in this spectral range, only the knowledge of
both reflectance and transmittance provides full information.
To discuss the different excitations a plot of
$\sigma (\hbar \omega )$ is most
suitable. There, the exponentially decreasing high energy
tail of the highest fundamental phonon mode is observed up to 0.15~eV.\@
Several smaller
structures due to absorption by multi-phonons are superimposed on it.
The main absorption features in the mid-infrared region are of magnonic
origin. The excitation of single-magnons  of the optical
branch is observed at $E_{\text{op}}$~=~178.0~meV.
The two peaks marked $E_{\text{2a}}$ and $E_{\text{2b}}$
both belong to the two-magnon spectrum, as will be discussed below.  The broad
high
energy tail of the spectrum is caused by higher multi-magnons.
Finally, the steep increase of conductivity above 1.3~eV is due
to the onset of intrinsic absorption, the excitation of carriers
across the charge transfer gap.

In order to interpret this magnon spectrum, we use linear spin wave theory
(LSW) to
gain the excitation spectrum of localized spins on a bilayered square lattice.
A Heisenberg
Hamiltonian accounts for these low energy excitations for zero doping:
\begin{equation}
H=\,J\!\sum_{a=1,2}\sum_{<i,j>}{\bf S}_{a,i}{\bf S}_{a,j}
                          +J_{12}\sum_{i}{\bf S}_{1,i}{\bf S}_{2,i}
\label{1}
\end{equation}
where $i$ and $j$ label nearest neighbor sites in a two-dimensional square
lattice
and $a\in \{1,2\}$ labels the two different planes in a single bilayer. Each
bond is
counted once. Generally it is found that LSW supplies quantitatively satisfying
results
for the N\'eel ground state at low temperatures\cite{chakra,manou,BK}.
The (classical) N\'eel ground state ${\bf S}_{{1\atop 2},\,i}\! =\!\pm (-1)^i\,
S\,(1,0,0)$
is stabilized by a finite bilayer coupling $J_{12}$\cite{bonesteel}
($S\!=\!1/2$). Spin-orbit
effects are relatively small\cite{bonesteel} and were neglected in Eq.\
(\ref{1}).
However, the finite spin-orbit coupling is needed to couple the external
electric
field to a single magnon, and further to make two-magnon absorption possible
for the considered crystal symmetry\cite{moriya}.

Due to the finite bilayer coupling, the classical one-magnon spectrum splits
into
acoustic and optical branches\cite{correct}:
\begin{equation}
\hbar\,\omega_{\text{op/ac}}(\bbox{k})=S J \sqrt{z^2-\tau_{\bbox{k}}^2
+2\,(J_{12}/J)\cdot
(z\pm \tau_{\bbox{k}})}
\label{3}
\end{equation}
where $z$=4 is the coordination number in a plane, and
$\tau_{\bbox{k}}=2\,(\cos(k_x a)+\cos(k_y a))$. $\bbox{k}$ is within the
magnetic Brillouin zone,
and $a$ is the lattice constant. Absorption experiments probe $\bbox{k}\!=\!0$
with energy gap
\begin{equation}
E_{\text{op}}\!\equiv\!\hbar\,\omega_{\text{op}}(\bbox{k}\!=\!0)=2(2S)\sqrt{J_{12}\,J}
\label{Eop}
\end{equation}
for the optical branch.
The acoustic mode splits, due to the small spin-orbit coupling\cite{bonesteel}.
The gapped out-of-plane mode is indeed observed
in neutron-scattering experiments at
$E_{\text{ac}}(\bbox{k}\!=\!0 )\!\approx\!4.5\!$~meV\cite{vettier}. The
splitting
of the optical branch at $\bbox{k}\!=\!0$ may be estimated to be
$\Delta E_{\text{op}}\!\approx\!
E_{\text{ac}}^{\,2}/2E_{\text{op}}\!\approx\!1/18\!$~meV,
a scale too small in comparison to the width of the optical magnon peak of
about $1\!$~meV
to be resolved in our experiment.

The two-magnon absorption is calculated with a coupling Hamiltonian of the form
\begin{equation}
H_1=D\sum_{a,b}\sum_{<i,j>}{\bf E}\cdot\left[({\bf S}_{a,i}\times{\bf S}_{b,j})
    \times\bbox{\pi}_{a,i\,;\, b,j}\right]
   \nonumber
\end{equation}
where $\bbox{\pi}_{a,i\,;\, b,j}$ points in the direction of the vector joining
the pair $\langle a,i\,;\, b,j\rangle$ and $\bf E$ is the electric field
vector\cite{elliott}.
This is the only coupling allowed by crystal symmetry
for a nearest neighbor two-magnon generation\cite{comment1}. $D\bbox{\pi}$ is
found from
a perturbation series in the long-wavelength electron-photon and
spin-orbit interaction\cite{moriya}. Since we restrict ourselves to $\bbox{k}
\!\! \parallel $c,
the two-magnon coupling is proportional to
$E^y\,\pi_{1,i\, ;
\,2,i}^z\cdot(S_{1,i}^{-}S_{2,i}^{+}\,-\,S_{1,i}^{+}S_{2,i}^{-})$ which creates
a
singlet pair of magnons on adjacent planes.

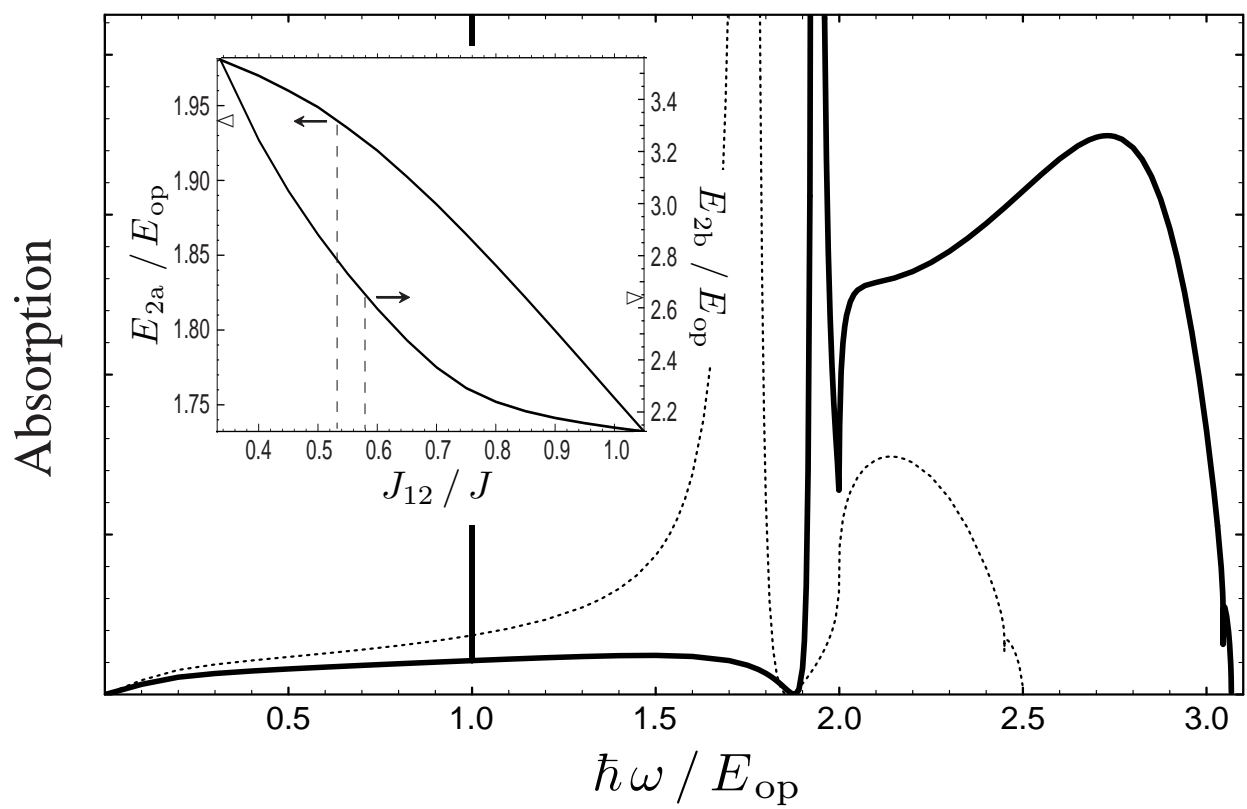
\begin{figure}
  \caption{Calculated two-magnon absorption for $J_{12}/J\!=\!0.55$
  (solid line) and 1.0 (dashed line) with $\protect{\bbox{k}} \! \!\parallel
$c, $T\!=\!0$.
  Inset: Peak positions of the lower ($E_{\text{2a}}$) and upper
($E_{\text{2b}}$)
  two-magnon resonance in units of $E_{\text{op}}$. Arrows indicate the
experimental values.}
 \label{fig2}
\end{figure}

Absorption is not just determined by  the convoluted density of states (DOS) of
two magnons
but also by the (local) interaction between the two magnons\cite{elliott}.
Therefore,
absorption is not just characterized by a step-like increase at the optical
two-magnon
edge, $2E_{\text{op}}$, and a diverging DOS at the upper band edge (as for
$J_{12}\!=\!0$).
Rather, the local interaction will reduce the frustration produced by the two
spinflips
and allow two optical magnons to form a nearly bound state below
$2E_{\text{op}}$.
Due to the admixture of acoustic magnons the bound state shows up
as a resonance (see Fig.\ \ref{fig2}). A second broader resonance close to the
band
edge replaces the diverging two-magnon DOS for $J_{12}\!>\!0$, similar to the
well-known
two-magnon peak in $B_{1g}$ Raman scattering. Quantum fluctuations certainly
broaden
the two-magnon peaks.

The exact positions of both peaks depend upon the ratio
$J_{12}/J$, as is displayed in the inset of Fig.\ \ref{fig2}. Comparison with
experiment
yields a value of $0.55\pm \,0.05$ for this ratio (0.53 for $E_{\text{2a}}$ and
0.58 for $E_{\text{2b}})$. Hence both resonances confirm the interpretation of
the peak
at $E_{\text{op}}$ as a single magnon process. The calculated minimum at
2$\cdot E_{\text{op}}\!=\!356.0\!$~meV
is also present in our measurements (see inset of Fig.\ \ref{fig1}). With
$E_{\text{op}}\!=\!178\!$~meV
and Eq.\ (\ref{Eop}) we obtain $J\approx\!~\!120\!$~meV and $J_{12}\!\approx
66\!$~meV.\@
For comparison, the calculated absorption spectrum for $J_{12}/J\!=\!1$ is
given by the
dashed line in Fig.\ \ref{fig2} which demonstrates that the shape has changed
qualitatively,
quite distinct from the measured spectrum. Actually, the resonance at
$E_{\text{2b}}$
is lost for $J_{12}/J\gtrsim 0.6$\cite{comment2} and is replaced by a broad
hump.

The amazingly high value of $J_{12}$ implies that the bilayer coupling may not
be
neglected in the interpretation of any experiment with YBa$_2$Cu$_3$O$_{6}$.
The slope of the magnon dispersion as measured in neutron scattering is
$\sqrt{2}\, 2S J\sqrt{1\!+\!{1\over4}
J_{12}/J}\!\simeq\!\sqrt{2}\,(J\!+\!J_{12}/8)$, but
$J_{12}/8$ is still a small correction within experimental uncertainty.
Raman-scattering results should
depend on $J_{12}$ more visibly:
The calculated zone boundary two-magnon peak in $B_{1g}$ geometry is located at
approximately
$E_{B_{1g}}\!\simeq2.7 J+J_{12}$ in LSW (Fig.\ \ref{fig3}). That is, the
excitation is a nearly local two-magnon state in a single plane with energy
$(2Sz\!-\!1)J + 2S\,J_{12}$ (cf.\cite{elliott}), and delocalization reduces the
frustration
in the plane. This calculated
peak position differs from that of recent experiments\cite{sugai} by about
$+15$\%.
But, the standard  Fleury-Loudon scattering Hamiltonian\cite{fleury}
$H_{\text{FL}}$,
which describes the interaction of spin pairs with light through a
spin-exchange process,
has to be drastically corrected for the high incident energies\cite{chubukov}.
These
are always above the charge transfer gap. Though, we only expect
a minor shift of the two-magnon peak position.

We propose that a major shift may
result from coupling to phonons: It was shown that the unexpected large width
$\Delta E_{B_{1g}}$ of the Raman peak may be explained by such a coupling which
is effective {\it only} for {\it zone boundary magnons}\cite{knoll,saenger}.
This mechanism
may be responsible for a shift of the $B _{1g}$ Raman peak of the order of
$(\Delta E_{B_{1g}}/E_{B_{1g}})^2$ but will effect the absorption peaks of
Fig.\ \ref{fig2}
only little.

Moreover, in $A_{1g}$ scattering geometry, which emphasizes the zone center, a
broad
resonance is observed at $E_{A_{1g}}\!\approx\!350\!$~meV\cite{sulewski}.
We expect this zone center peak not to be
shifted by phonons. Indeed, LSW reproduces such a broad feature with a
maximum close to the measured $E_{A_{1g}}$ (inset of Fig.\ \ref{fig3}). To
ensure that
$A_{1g}$-scattering vanishes for $J_{12}=0$, fluctuations of the longitudinal
component
$S^x_i S^x_j$ have to be added to the response functions\cite{elliott}
(inset of Fig.\ \ref{fig3}, dashed line).
Although a consistent calculation including these fluctuations is beyond LSW
the
trend of such a correction is clearly seen: the asymmetry of the $A_{1g}$
resonance
is more pronounced, and the intensity is reduced. However, even if the shape is
reproduced
correctly, other processes, such as next nearest neighbor contributions to
$H_{\text{FL}}$,
may be as important. Besides, also single layered cuprates exhibit a
$A_{1g}$ resonance which
does not result from $H_{\text{FL}}$.

In conclusion, we presented the low conductivity mid-infrared spectrum of
YBa$_2$Cu$_3$O$_{6.0}$, which is dominated by absorption due to magnons.
By comparison with linear spin wave theory we could determine the ratio
$J_{12}/J\!=\!0.55\pm 0.05$ and hence $J\!\approx\!120\!$~meV and
$J_{12}\!\approx\!66\!$~meV.\@

We are indebted to P.~W\"olfle for many helpful discussions and his
encouragement,
and we also acknowledge useful discussions with A.~Rosch.
This work was supported by the Commission of the EC under
contract no.\ CI 1-0526-M(CD) and by the High-$T_c$ program of
Baden-W\"urttemberg. It was supported in part by the DFG (M.G., H.P.G.) and
by SFB 195 (T.K.).

\begin{figure}
  \caption{Calculated Raman scattering in $B_{1g}$ geometry
   for $J_{12}/J\!=\!0.55$, $T\!=\!0$. The Raman shift $\hbar\Delta\omega$ is
displayed
   in units of $E_{\text{op}}$ to compare the peak position to the position of
the absorption
   resonances in Fig. 2.
   Inset: $A_{1g}$ geometry  in  LSW (solid line), with corrections from
longitudinal spin
   fluctuations (dashed line, scaled by a factor 3.5), both for
$J_{12}/J\!=\!0.55$.}
 \label{fig3}
\end{figure}


\begin{references}
\bibitem{comment2}i.e., the real part of the denominator of the response
function
has no zero at $E_{\text{2b}}$ for $J_{12}/J\gtrsim$0.6.
\end{references}
\end{document}